\documentclass[onecolumn,authoryear]{els-mrw} 

\usepackage{amsmath,amssymb,amsfonts,amsthm,makeidx,graphicx}
\usepackage{txfonts}
\usepackage{helvet}

\usepackage{silence}
\usepackage{hyperref}
\hypersetup{hidelinks}

\usepackage{journal_macros}

\makeatletter
\newcommand{\ion}[2]{\mbox{#1\check@mathfonts\fontsize\sf@size\z@\selectfont #2}}
\makeatother

\begin{document}

\chapter{Wolf-Rayet Colliding Wind Binaries}\label{chapter}

\author[1,2]{Ryan~M.T.~White}%
\author[3]{Peter Tuthill}%

\address[1]{\orgname{School of Mathematical and Physical Sciences}, \orgdiv{Macquarie University}, \orgaddress{N.S.W. 2109, Australia}}
\address[2]{\orgname{School of Mathematics and Physics}, \orgdiv{The University of Queensland}, \orgaddress{St Lucia, QLD, 4072, Australia}}
\address[3]{\orgname{Sydney Institute for Astronomy}, \orgdiv{School of Physics}, \orgaddress{The University of Sydney, N.S.W. 2006, Australia}}

\articletag{Chapter Article tagline: update of previous edition,, reprint..}

\maketitle

\begin{glossary}[Glossary]
    \term{Central engine} refers to the binary stars at the heart of a colliding wind binary
    
    \term{Colliding wind nebulae} are dust structures created by shocked winds of two stars, forming a spiral wrapped by the orbital motion of the stars. These are sometimes referred to as \textbf{pinwheel nebulae}.

    \term{Eddington limit} is the point at which the radiation pressure from a luminous source cancels out the radially inwards gravitational attraction. 
    
    \term{Hydrogen envelope} refers to the outermost hydrogen shell of a star, where beneath this are heavier elements such as helium, carbon, etc.

    \term{Wind collision region} refers to the turbulent bow shock structure where the winds of two stars are directly colliding.
    
\end{glossary}

\begin{glossary}[Nomenclature]
    \begin{tabular}{@{}lp{34pc}@{}}
        BSG & Blue Supergiant (star)\\
        CWB & Colliding Wind Binary \\
        GRB/LGRB & Gamma Ray Burst and Long GRB respectively \\
        JWST & James Webb Space Telescope \\
        LBV & Luminous Blue Variable (star) \\
        OB & O or B type star \\
        RSG & Red Supergiant (star)\\
        SN/SNe & Supernova and Supernovae (plural form) respectively\\
        VLT & Very Large Telescope \\
        WC/N/O & Wolf-Rayet star of Carbon, Nitrogen, or Oxygen type respectively \\
        WCR & Wind Collision Region\\
        WNh & WN star with hydrogen spectra \\
        WR & Wolf-Rayet (star)\\
    \end{tabular}
\end{glossary}

\begin{abstract}[Abstract]
Wolf-Rayet stars embody the final stable phase of the most massive stars immediately before their evolution is terminated in a supernova explosion.
They are responsible for some of the most extreme and energetic phenomena in stellar physics, driving fast and dense stellar winds that are powered by extraordinarily high mass-loss rates arising from their near Eddington limit luminosity. 
When found in binary systems comprised of two hot wind-driving components, a colliding wind binary (CWB) is formed, manifesting dramatic observational signatures from the radio to X-rays. 
Among the wealth of rare and exotic phenomenology associated with CWBs, perhaps the most unexpected is the production of copious amounts of warm dust.
A necessary condition seems to be one binary component being a carbon-rich WR star -- providing favorable chemistry for dust nucleation from the wind -- however a detailed understanding of the physics underlying this phenomenon has not been established.
\end{abstract}

\begin{BoxTypeA}[chap1:box1]{Key Points}
    \begin{itemize}
        \item The winds of each star in binary systems collide to form bow shocks. These shocks encode the physics of each stellar wind, and are visible across the electromagnetic spectrum through both thermal and non-thermal emission.   

        \item The systems hosting a Wolf-Rayet star are the sites of the most extreme colliding wind binaries; these Wolf-Rayet stars typically have orders of magnitude higher mass loss rates which affect the wind-wind shock physics. 

        \item Some Wolf-Rayet colliding wind binaries -- where the conditions at the shock are just right -- are prolific dust producers. This dust traces the orbital motion of the shock with the binary stars, presenting itself as spiral nebulae that are bright in the infrared. 

        \item The observational characteristics of the Wolf-Rayet colliding wind binaries are tied to orbital phase, and can result in transient peaks in X-ray/infrared/radio emission. Continued monitoring of these systems reveals their orbits, wind physics, and stellar parameters. 

        \item Modelling of these systems is undertaken in a variety of ways depending on the wavelength regime and type of data. Photometric, spectroscopic, hydrodynamical, and geometric modelling are all commonly used to investigate colliding wind binaries and the Wolf-Rayet and massive stars at their heart.

    \end{itemize}
\end{BoxTypeA}


\section{Introduction} \label{sec:intro}

The understanding of massive stars in our Galaxy is foundational to our understanding of the wider Universe. Massive stars are born from the densest regions within clouds of molecular gas and begin shaping their environment and their galaxy from the outset. These stars evolve very quickly -- on the order of a few to tens of millions of years -- and produce an abundance of metals as they do so. As they live, massive stars eject material into the interstellar medium through their strong stellar winds and through episodic eruptions. As they die, massive stars enrich the chemical content of their galaxy through powerful supernova explosions. These two processes from massive stars, stellar mass loss and supernovae, are responsible for a large fraction of the metal content in the Universe and in particular they dominate production at early epochs. For example Wolf-Rayet (WR) stage of stellar evolution contributes the majority of early carbon, nitrogen, and oxygen without which our Solar System and life as we know it could not have formed. 

These Wolf-Rayet stars embody the rare and extreme end-of-life phases of the most massive stars that have shed their hydrogen envelopes through binary interactions and/or the strongest known stellar winds. Despite their intrinsic rarity -- only 1 in every billion stars within the Galaxy are of WR-type -- these stars exert a profound influence on the Milky Way, for example in their role as progenitors of the most energetic core-collapse supernovae: Type Ic supernova. These supernovae have been associated with long duration gamma-ray bursts: events so luminous that they are visible across the observable universe. Galactic gamma ray bursts are thought to be rare -- a fortuitous outcome for life on Earth as a the impacts from an event unluckily beamed along our line of sight could be significant. Exactly how WR stars in the local Universe may produce a gamma-ray burst is yet to be conclusively understood as the mechanism should be disfavoured by higher metallicity at contemporary epochs. It is hypothesised that binary interactions may play a key role, imparting or preserving angular momentum of the progenitor WR star sufficient to break spherical symmetry of the supernova, enabling the gamma ray burst to take place. Since a majority of massive stars are formed in binary or multiple systems, WR binaries are therefore of particular interest.

With no confirmed direct in-situ observations of a Wolf-Rayet precursor undergoing a supernova explosion yet recorded, the current population of WR stars must be closely studied to advance our understanding of the final moments of massive stars. Of particular interest are the colliding wind binaries (CWBs) that host WR stars. When a WR star and a sufficiently massive companion are in a binary system, a wind collision region (WCR) is formed yielding observational signatures that can include non-thermal radio emission from the shock, and in rare instances, the formation of carbon-rich dust nucleated in association with the collision interface between the components' stellar winds. Such dust can be bright in the infrared streaming into circumstellar structures inflated by the stellar wind. These dusty nebulae carry structural elements engraved by the orbital motion of the colliding-wind binary itself, forming an intricate spiral structure encoding a wealth of key astrophysical quantities. For the simplest case when the orbital plane lies close to the plane of the sky, the dust nebula forms an Archimedean spiral, or a so-called `pinwheel nebula' as a simple consequence of a rotating insertion (orbit) of matter (dust) into a spherically expanding outflow (wind). This spiral structure is sensitive to the orbital and wind parameters of both system components, encoding the underlying physics of the CWB, and by extension the properties of the hots stars that comprise the engine driving the system.

\section{Origins and Basic Properties}

\subsection{Formation of Wolf-Rayet stars}
Massive stars -- for this discussion those of order 10 to 100$M_\odot$ -- form in giant molecular clouds along with their lower mass brethren \citep{deWit2005A&A, Tan2014prpl.conf}. The initial masses of stars formed within these molecular clouds follow an initial mass function (IMF) such that low mass stars are formed in much higher number than high mass stars \citep{Miller1979ApJS, Kroupa2001MNRAS}. Despite their intrinsic rarity, massive stars' vastly greater intrinsic energy output, both in the form of luminosity and high momentum winds, results in them being the dominant force in shaping their parent molecular clouds once star formation begins \citep{Motte2018ARA&A}. The profound impacts even extend to galactic scales, influencing the appearance of their host galaxy \citep{Cameron2024MNRAS}, for example ionising the gas in \ion{H}{II} regions.
Because massive stars naturally form in the densest central regions of molecular clouds (and in addition to their correspondingly higher gravitational influence), there is a strong preference for binarity/multiplicity in massive star systems, and with companion stars also tending towards comparably high masses \citep{Sana2011IAUS, Sana2017IAUS, Offner2023ASPC}. 

A few million years after their birth, the hydrogen in the convective cores of very massive stars can be exhausted so that the star leaves the main sequence, forming heavier elements via the triple-$\alpha$ process \citep{Salpeter1952ApJ, Woosley2002RvMP, CarrollOstlie2017imas.book}. During a phase of core/shell helium or carbon-oxygen burning, the stars then usually evolve into supergiant stars. Depending on the initial mass of the star, this could mean a red or blue supergiant (RSG and BSG respectively), or a luminous blue variable (LBV) star for the more massive stars \citep{Groh2013A&A}. Whether through one or a combination of mass loss mechanisms such as stellar winds and/or binary mass transfer, the highest mass ($\gtrsim 20M_\odot$) stars shed their outer hydrogen envelopes leaving an exposed helium star or Wolf-Rayet star \citep[WR;][]{Paczynski1967AcA, Conti1975MSRSL, Crowther2007ARA&A}. The exact progression of envelope/mass loss is a topic of active research, and depends very sensitively on the mass, metallicity, and rotation of the star among other variables \citep{Meynet2011BSRSL, Sander2020MNRAS, Josiek2024arXiv}; the most massive stars ($>25M_\odot$) may skip the supergiant phase altogether and evolve off the main sequence directly into a WR star \citep{Crowther2007ARA&A, Groh2013A&A}.

\subsection{The Wolf-Rayet Spectrum and Stellar Wind}
\begin{figure}
    \centering
    \includegraphics[width=\linewidth]{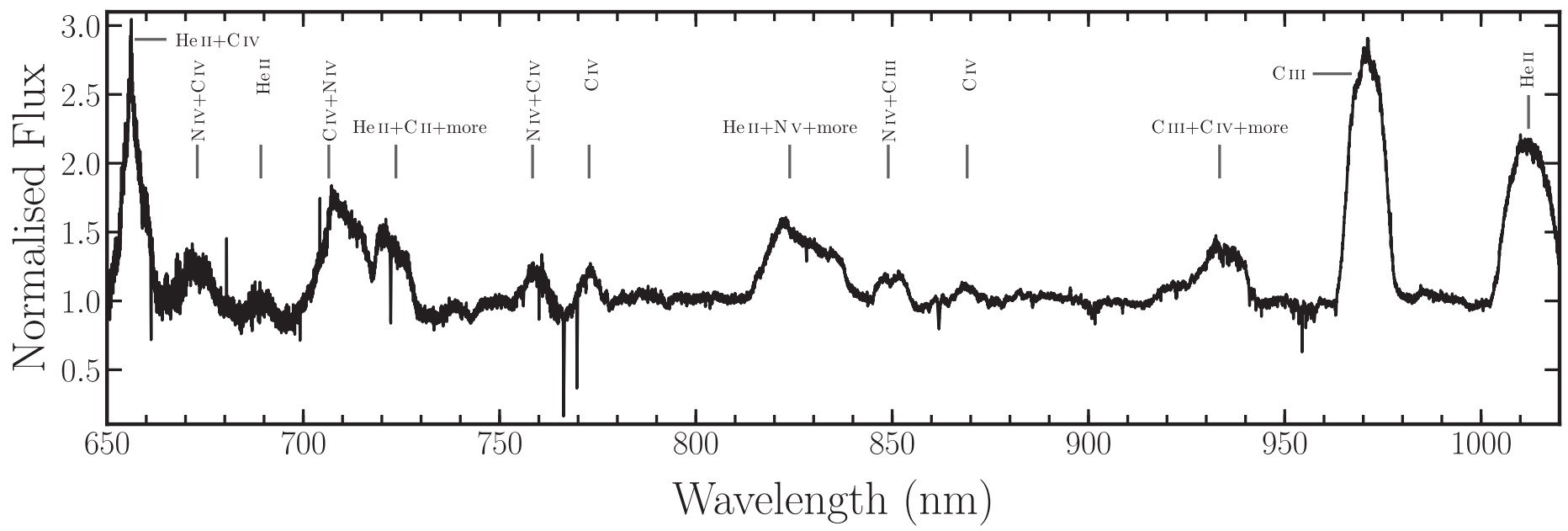}
    \caption{The spectrum of the WR+WR system Apep shows emission lines of both WC and WN Wolf-Rayet species \citep{Callingham2020MNRAS}. In particular, the WC star is mainly identified through \ion{C}{III} and \ion{C}{IV} lines while the WN star through the \ion{He}{II} and N lines. }
    \label{fig:apep_spectrum}
\end{figure}
Wolf-Rayet stars are characterised by extremely strong, dense stellar winds driven by  emission lines of helium, carbon, nitrogen, and/or oxygen \citep[][and Figure~\ref{fig:apep_spectrum}]{Wolf1867CRAS, Hamann2006A&A, Crowther2007ARA&A}. Despite the hydrogen envelope shedding and high mass loss rates, WRs remain very massive at $\gtrsim 10M_\odot$, depending on their spectral type. Within the Wolf-Rayet classification, there are 3 main subclassifications labelled for carbon, nitrogen, and oxygen: the WN type (owing to strong He and N lines), the WC type (owing to an absence of N and prominence of C), and the WO type (an absence of N and strong presence of O) \citep{Gamow1943ApJ, Crowther2007ARA&A}. Within each of these subclassifications there exist subtypes that describe the relative strength of individual emission lines (WN2-11, WC4-9, and WO1-4) and hence more specific chemical and physical properties of the stars. The relationship between the different classes of WR stars has long been understood to be an evolutionary progression from one type to another \citep[depending on the initial mass of the star;][]{Conti1975MSRSL, Abbott1987ARA&A}, although this has been recently challenged \citep[][suggesting that the subtype is mass-dependent]{McClelland2016MNRAS} indicating that even the fundamental properties of these stars are still a topic of active research. 

In the last 30 years, there has been interest in a peculiar class of WR star: the WNh type. Wolf-Rayet stars of the WNh subclassification are categorically disparate from the classical WRs of C/N/O type; while they do display strong helium and nitrogen spectral lines characteristic of WN stars, they also exhibit hydrogen lines (hence the appended ``h") which suggests either a transitory stage of evolution or a different phenomenology altogether \citep{Shenar2024arXiv}. The $60-100M_\odot$ mass of these stars -- a factor of a few higher than typical WC/WN stars -- cements these as massive stars still undergoing core-hydrogen burning, although perhaps near the terminal-age main sequence \citep{de-Koter1997ApJ, Smith2008ApJ, Shenar2024arXiv}. Still, their mass loss rates are many times higher than ordinary main sequence O stars and their winds are expected to be radiatively driven, making all but their chemistry similar to classical Wolf-Rayets at first glance.

\begin{figure}
    \centering
    \includegraphics[width=0.8\linewidth]{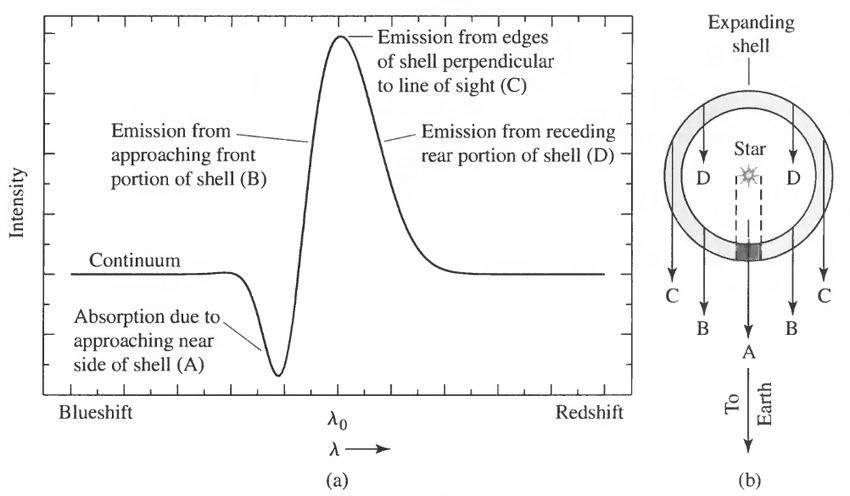}
    \caption{Stellar spectra often show a P~Cygni profile: a distinct absorption feature that is blueshifted with respect to a broader emission line [figure (a)]. This comes about due to material expanding spherically away from a central star, so that cold distant material with the highest apparent speed (directly approaching the observer) absorbs emission from interior regions, while light emitted by material streaming in other directions carries a different Doppler shift and is therefore not in resonance with the absorbing gas along the line of sight [figure (b)]. 
    Figure from \citet{CarrollOstlie2017imas.book}. }
    \label{fig:p-cygni}
\end{figure}

Despite the unknowns in their evolutionary progression, there is much about the phenomenology of Wolf-Rayet stars is believed to be well established. Arguably their most noteworthy feature is their uniquely fast, dense and powerful stellar winds. The properties of these winds have been established by multiple observational probes that reveal both intrinsic \citep[e.g. line broadening in spectra;][]{Milne1926MNRAS} and extrinsic (in the interaction with the WR surrounds) phenomenology. 

The most robust and well established indicator of a WR's wind velocity can be found from its spectrum. Observing individual emission lines in the spectrum of a wind-driving star can show a P~Cygni profile (Figure~\ref{fig:p-cygni}), wherein the emission line, doppler broadened by the spherical expansion of the luminous wind, has an absorption feature engraved at a wavelength corresponding to resonance with atoms streaming along the line of sight towards the observer and therefore blueshifted at the wind terminal velocity. Analysis of the shape of this line profile can yield highly precise wind speeds \citep{Willis1982MNRAS, Crowther2007ARA&A, Callingham2020MNRAS, Williams2021MNRAS}. \citet{Zavala2022MNRAS} also showed that by taking several spectra at different regions of a WR nebula, a three dimensional geometric model of the nebula can be produced that encodes information not only about the wind speed, but constrains wind direction too.

Interactions of the Wolf-Rayet wind with the surrounding environment, such as the creation of nebulae or cavities in parent molecular clouds, can yield strong constraints on wind physics.
In addition to those from fast WR winds, such strong interactions also arise from several classes of star exhibiting high mass loss rates, with WR phenomenology comparable only to LBVs and RSGs. It is understood that this mass loss is primarily driven by the radiation pressure acting upon metallic emission lines \citep{Castor1975ApJ, Crowther2007ARA&A} as the WR stars themselves lie near the Eddington limit \citep{Grafener2008A&A, Grafener2011A&A}. Mass loss rates for WR stars are typically of order $10^{-5}M_\odot$\,yr$^{-1}$ \citep{Barlow1981MNRAS, Crowther2007ARA&A}, and so this coupled with the high wind velocity means that WR winds carry exceptionally high momentum. 

Over the course of the WR phase, the stellar winds of the stars can form bright nebulae (see: NGC~2359, NGC~6888, Sh~2-308) fuelled by their mass and enriched chemistry; studying the multi-epoch expansion of such nebulae delivers distances to these systems if the wind velocity is well constrained (or vice versa). One such case where nebula expansion has been observed is in the wind nebula M1-67 around WR\,124 \citep{Marchenko2010ApJ}, which allowed for an independent distance estimate to the system; this has also been done with the WR PNe BD+30$^\circ$3639 \citep{Schonberner2018A&A}. As a secondary direct imaging probe of WR winds, WR stars have been observed forming cavities in the molecular material of their natal environment via ablation from their stellar winds \citep[][see also \citealp{Dale2013MNRAS} for simulations of this]{Baug2019ApJ}. This same process has been seen forming shocks when the wind interacts with the surrounding gas, sometimes triggering more star formation \citep{Cichowolski2015MNRAS}.

At some scales, mass within the stellar winds is likely inhomogeneous: clumpy dust nucleation processes are believed to feature across many astrophysical theatres. Although WR stars drive mostly steady winds, there is an observed small-scale stochasticity in the winds that manifests as multi-epoch line variation \citep{Michaux2014MNRAS, Chene2020ApJ}; hence, the WR stellar wind can be chemically inhomogeneous. On a related note, circumstellar dust is produced around many WR stars, especially those of WC subclassification \citep{Allen1972A&A, Gehrz1974ApJ, Lau2021ApJ}. How delicate dust molecules are formed around WR stars -- some of the hottest and most ionising astrophysical environments in existence -- is still a topic of active research. The clumpy wind is one proposed avenue for dust formation, where the high opacity and density within clumps provides a sort of shielding from the environment where molecules can form \citep{Cherchneff2000A&A}. We observe orders of magnitude more efficient dust production around binary stars containing a WR, where turbulent wind compression occurs in the shocked region between the two close massive stars, providing a nurturing environment for dust creation and survival \citep{Soulain2023MNRAS}.

\subsection{Binaries and Colliding Winds}

While it has been established that massive stars preferentially form in binary or multiple systems within the Galaxy \citep{Offner2023ASPC}, modellers also find an apparent link between the rotation rate of massive stars and their binary status. Such findings begin to fill in a broader canvas of how Wolf-Rayet stars fit within the picture of binary/multiple systems, and what this means for their evolution and observation.

Massive O stars on the main sequence are most commonly formed within binary or multiple systems, and with a predilection for companions to also be massive of O or B (sometimes abbreviated to ``OB'') spectral type \citep{Shara2022MNRAS, Offner2023ASPC}. Naturally, as the primary stars in these systems evolve off the main sequence into BSG/RSG or WR stars, they tend to keep their companions. Eventually, the primary star is expected to undergo a supernova explosion in which case the orbit may be disrupted causing the secondary to be ejected as a `runaway star' with high velocity \citep[typically of order the orbital velocity at the time of supernova, up to $\sim120$\,km\,s$^{-1}$;][]{Eldridge2011MNRAS}. This means that although surveys identify isolated O/B/WR stars, these may have originated within a now-disrupted binary system \citep{Schootemeijer2024arXiv}. Regardless, the Galactic WR binarity fraction is reported as $\sim40$\% \citep{van_der_Hucht2001NewAR, DeMarco2017PASA} with surveys of the (lower metallicity) Large and Small Magellanic Clouds (LMC and SMC, $Z\sim0.5Z_\odot$ and $Z\sim0.2Z_\odot$ respectively) suggest similar values. Indications are that binarity is not a strong function of either metallicity or redshift \citep{Foellmi2003MNRAS, Shenar2020A&A, Schootemeijer2024arXiv}.

Beyond the traits encompassed in the classification (two or more stars per system at least one of which is a Wolf-Rayet), the WR multiple systems seem to conform to few common rules. Companions have been found in both very close and very wide orbits, and interestingly there appears to be at least some correlation between orbital period and WR type: WN stars are more commonly found in close binaries, and WC in wide binaries \citep{Dsilva2020A&A, Dsilva2022A&A, Dsilva2023A&A}. Also, while WR companions are usually similarly (or more) massive OB type stars, there are several confirmed and tentative detections of unusual companion types. Of particular note is the possible detection of a lower mass F type main sequence companion to WR\,113, making the system a triple with its known WC8+O8 components \citep{Shara2022MNRAS}. In the same study the authors propose a detection of a wide orbit ($\sim 1800$\,au) WN3-4 companion to the WN7 star WR\,120, making it one of the few candidate WR+WR systems. Although not evolved stars, the WR\,20a system has been observed to contain a WN6h+WN6h binary at its centre in an extremely short $\sim 4$\,day orbit \citep{Bonanos2004ApJ, Rauw2005A&A, Olivier2022arXiv}. WR\,48a has been proposed as a candidate WC8+WN8h system on the basis of its spectra \citep{Zhekov2014MNRAS}, although this is contested in favour of an O star companion classification \citep{Williams2012MNRAS}. 

Relatively recently discovered, the spectacular Apep (WR\,70-16) system has been confirmed as a WC8+WN4-6 binary due to the superposition of specific spectral lines characteristic of those two classes of WR star, together with a clear wind speed discrepancy in those lines \citep{Callingham2020MNRAS}. At present, Apep is the only confirmed WR+WR system composed of two evolved WR stars. The evolutionary history delivering two Wolf-Rayet stars in the same system at the same time is, like much of Wolf-Rayet evolution in general, an open question and made still more puzzling by the disparate spectral classes (WC and WN). Early studies even questioned the observability of these systems \citep{Vrancken1991A&A}, although technique and instrumentation development has provided clear success in detection of WR binaries. With a lack of population synthesis studies showing a clear WR+WR formation channel, one possible pathway forward may lie in the study of less evolved massive stars \citep[e.g. in the WN6h+O3/WN7 system R145;][]{Shenar2017A&A}.

Given that Wolf-Rayet stars are known for their powerful winds and high binarity fraction, it naturally arises that we should expect some interaction phenomenology at the collision interface between the winds of the WR star and its (usually massive) companion \citep{Cherepashchuk1976SvAL, Prilutskii1976SvA}. Indeed, there is a modest population of WR+O/B/WR binaries whose dense winds collide to produce dust nebulae that expand outwards from the system. We illustrate the diversity of such systems in Figure~\ref{fig:CWBs}, revealing that even the same physics can produce vastly different observed nebula geometry.

\section{Phenomenology}

\begin{figure}
    \centering
    \includegraphics[width=0.49\linewidth]{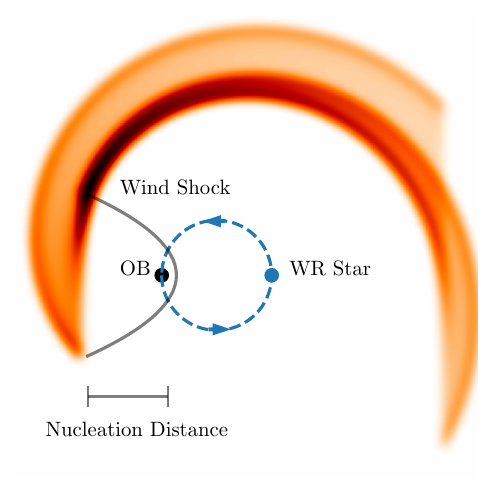}
    \includegraphics[width=0.49\linewidth]{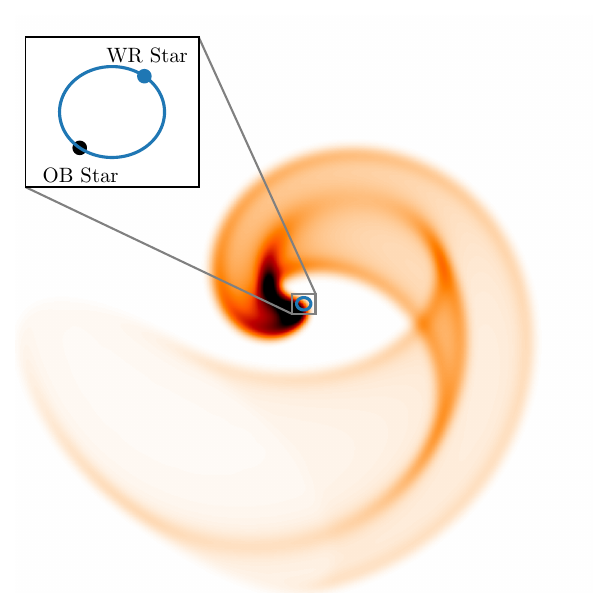}
    \caption{\emph{Left:} A top down view of a representative WR+OB colliding wind binary (not to scale). The WR stellar wind shocks the OB stellar wind, and dust is produced downstream (at the nucleation distance) once the shock has sufficiently cooled. As the two stars orbit (with the path defined by the dashed blue line), the produced dust nebula (nebulous orange region) wraps into a spiral. \emph{Right:} Similar to the left plot, but inclined to $i=30^\circ$ and at more realistic scale where the nebula is significantly larger than the orbit.}
    \label{fig:colliding_winds}
\end{figure}

\subsection{Infrared}

Among the earliest infrared observations of WR stars in the 1970s uncovered increased emission particularly in the continuum to longer wavelengths, a finding known as an “infrared excess” \citep{Allen1972A&A, Roche1984MNRAS}. As dust was not ordinarily expected in the hostile environs surrounding luminous hot blue stars, the infrared excesses were originally thought to arise from free-free emission \citep{Hackwell1976ApJ}. However, further observations following an infrared brightening in WR~140 revealed that the spectral energy distributions were, indeed, consistent with hot dust \citep{Williams1978MNRAS}. This established WR~140 as something of a touchstone for the whole field -- a status it has maintained for a half century. Because high gas densities and relatively low temperatures are considered necessary conditions for dust formation, the dust-producing WR stars posed a conundrum for astrophysics: how can copious circumstellar dust form in the immediate environment of stars with near relativistic winds, harsh UV radiation fields and extreme temperatures? 

Observers found that for the latest subtypes of the WC series -- classes WC7, WC8 and beyond -- the presence of dust became increasingly common, a finding attributed to the relative abundance of enriched material, particularly carbon, in the stellar wind.  
Addressing the riddle of the dust formation mechanism, the association of the periodic infrared brightening of WR\,140 \citep{Williams1978MNRAS, Williams1987QJRAS} with periastron passage of an eccentric binary precipitated the first major breakthrough in understanding the origins of the dust.
A model emerged in which the wind collision region between the two stars was able to create and harbour a dust nursery where wind compression enhanced the density, with radiative cooling further tipping the ambient conditions in favour of nucleation \citep{Williams1990MNRAS, Usov1991MNRAS}.
wind collision region and subsequent turbulent shock provides an ample environment for dust production \citep{Williams1990MNRAS, Usov1991MNRAS}. Further mechanisms might include mixing of chemical ingredients (such as Hydrogen) at the contact discontinuity and subsequent shielding from the ionising radiation in the turbulent conical shock boundary \citep{Hendrix2016MNRAS}. While details of the dust formation mechanism may remain uncertain, resulting infrared bright nebulae adorning a number of systems have generated considerable attention, yielding intricately structured forms whose geometry and periodic behaviour encode otherwise hidden orbital and wind properties of the system. 
% The spiral structure in these nebulae arise from the shock cone produced at the collision interface wrapping around from the orbital motion of the stars. 

The touchstone colliding-wind binary system WR\,140 is today regarded as the archetype of the episodic dust producers, while persistent dust emitter WR\,104 is the prototype of the so-called `pinwheel nebulae'. The latter system features a circular binary orbit resulting in little intrinsic variation in the properties of the wind-wind shock and as a consequence a continuously uniform dust production rate in contrast to the orbitally-modulated IR excess for WR~140. The simplest geometry arises where the binary orbital motion lies in the plane of the sky, producing an apparent perfect spiral of dust -- a structure that seems suited to the observed appearance of WR\,104 \citep{Tuthill1999Natur}. 
% With this in mind, we see so called `pinwheel nebulae' produced in binary systems where the components have roughly circular orbits and orbit roughly in the plane of the sky (e.g. in WR\,98a and WR\,104, the latter shown in Figure~\ref{fig:CWBs}).
The simplest form of spiral \citep[][originally described by Archimedes in approx. 225\,BCE]{Archimedes} has the radial extent of a point on the curve determined purely by its outwards velocity and angular coordinate,
\begin{equation}
    r = v \cdot \theta \label{eq:archimedean_spiral}
\end{equation}
Despite the simplicity of the model, for CWBs in a circular orbit constantly producing dust these Archimedean spirals have been found to trace the nebula geometry exceptionally well and dust position described by Equation~\ref{eq:archimedean_spiral} fits well to data from a number of systems. Where the central binary system exhibits significant ellipticity in its orbit, then the resulting spiral will deviate from the simplest family of shapes, and furthermore such system are usually also associated with orbital modulation in the rate of dust production. Although this can manifest as a relatively smooth variation in the infrared excess with orbital phase, it more often seems to result in a `turn on' and `turn off' in the spiral (see Apep, WR\,140 in Figure~\ref{fig:CWBs}). Clearly this speaks to some degree of criticality in the physics of dust production, with accumulating evidence that dust production will shut off relatively quickly both under conditions where the wind shock is too weak, and where it is too strong. Such extra complexities on the geometry imposed by eccentricity and orbital modulation of dust production rate, when also coupled with the range of random inclinations that systems exhibit relative to our line of sight, results in no two colliding wind nebulae appearing the same despite the consistent and relatively simple geometry of the formation mechanism. 

Although the dust nebulae produced by colliding wind binaries are orders of magnitude larger than the Solar System (see the scales in Figure~\ref{fig:CWBs}), their Galactic distance of $\gtrsim1$\,kpc requires that high angular resolution observations are needed to resolve structure. This, coupled with the fact that the resolved dust nebulae are most luminous in the infrared, means that direct imaging of these nebulae has been a relatively recent development coevolving with high angular resolution infrared astronomy. The recovery of direct imagery of these nebulae is an essential effort as we can precisely constrain many fundamental properties of the orbit and wind that cannot be recovered by spectral or photometric analyses (see Section~\ref{sec:infrared_modelling}).

Most of the earliest examples of full images for these systems, such as those for WR\,104 and WR\,98a, were recovered with aperture masking interferometry on the Keck~1 telescope \citep{Tuthill1999Natur, Monnier1999ApJ, Tuthill2000PASP}. This method allowed for fully diffraction-limited observation at sufficiently high angular resolution to recover complex asymmetric structures on the characteristic scales demanded by the system architecture itself ($\lesssim 100$\,mas). The intervening quarter century has seen an explosion of ground-based high angular resolution infrared astronomy, where direct images of colliding wind nebulae has become almost routine on large, 8~metre class telescopes \citep[][for WR\,112, 48a, 104/98a, and Apep respectively]{Marchenko2002ApJ, Marchenko2007ASPC, Tuthill2008ApJ, Callingham2019NatAs}, even for very distant systems near the galactic centre \citep{Tuthill2006Sci}. 

Most recent, the \emph{James Webb Space Telescope} (JWST) has begun observing colliding wind binaries. JWST has revealed dust structures that would otherwise be difficult or impossible to obtain from ground-based facilities due to sensitivity and resolution limits imposed by the atmosphere. This has allowed observation of cooler dust at a greater distances from the central CWB engine, thereby delivering simultaneous observation of dust generated over a number of successive orbital periods taking the form of a ladder structure of nested shells \citep{Lau2022NatAs}. Not only does this allow us to determine the behaviour of dust over $\sim$ hundreds of years (its cooling curve and lifetime, for example), this allows us to break degeneracies in our models and quantify fundamental orbital and wind parameters governing the CWB systems at very high precision.

\subsubsection{Photometry and Spectra}
\begin{figure}
    \centering
    \includegraphics[width=.5\columnwidth]{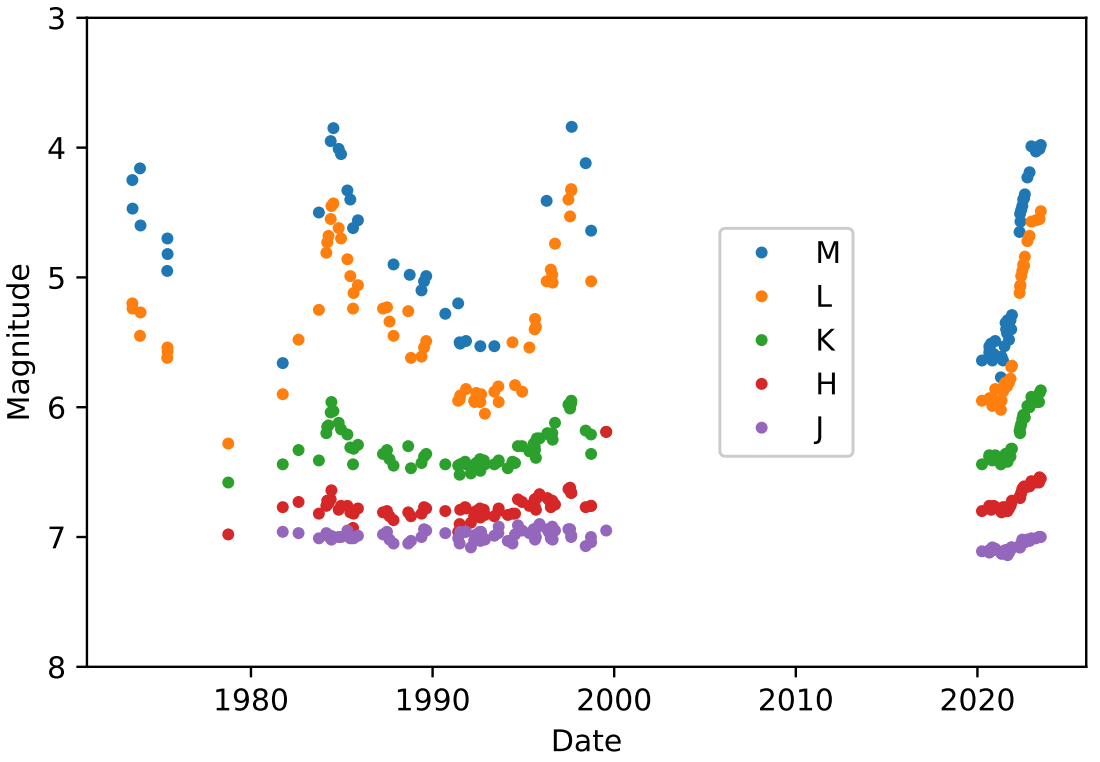}
    \caption[The $\sim$50 year light curve of WR\,137]{The long-term infrared light curve of WR\,137 shows episodic dust production in the periodic peaks. Shown here is the photometry across different infrared bands (descending band central wavelength in the legend), where the deeper infrared shows brighter total emission as a result of dust production. As the stars approach periastron every $\sim13$ years, dust production turns on and eventually turns off after the stars are again sufficiently far apart. Figure from \cite{Peatt2023ApJ}.}
    \label{fig:cwb_photometry}
\end{figure}

From earliest discoveries through to contemporary studies, direct photometric and spectral observation has been the workhorse tool for understanding colliding wind binaries. Systems exhibiting episodic dust production, especially those with highly eccentric orbits, yield clear photometric signatures due to their periodic peaks in infrared brightness (e.g. for WR\,137 in Figure~\ref{fig:cwb_photometry}). This is especially notable in systems such as WR\,140 ($P\sim7.9$ years) and WR\,48a ($P\sim32$ years) where mostly complete light curves have now been compiled since discovery \citep{Williams1987QJRAS, Williams2012MNRAS, Peatt2023ApJ, Richardson2024arXiv}. This is potentially the most efficient way to discover new CWBs: photometric monitoring can be obtained from sky survey data and only requires discrete snapshots distributed in time (perhaps even when only in the field of another target). Light curve analysis is still being used to discover new episodic dust producing CWBs \citep{Williams2019MNRAS} as well as being a useful tool to constrain properties such as the dust production rate when monitoring known systems. 

Among the earliest tools for the study of dust formation in Wolf-Rayets was the analysis of their spectra. Dust produced in these systems is relatively cool (of the order $\lesssim 1000$\,K), and so has peak blackbody emission in the infrared \citep[][for dust peaks around $10-15$\,$\mu$m in WR\,48a, 98a, 104, 112, and 118]{van_der_Hucht1996A&A}. Circumstellar dust absorbs higher energy starlight, re-emitting in the infrared where the overall spectral energy density therefore indicates a luminosity far in excess of dust-free stars \citep{Gehrz1974ApJ, Williams1978MNRAS}. Analysing the spectra of these systems remains a key tool in understanding their wind composition, and also their orbits \citep[see for example the spectral decomposition of WR stars in][]{Callingham2020MNRAS}.

A recent spectral analysis of a dust shell of WR\,140 suggested that there exists not just one infrared peak of carbon dust emission, but two: one dust species of $T\sim1000$\,K and grain size $\sim1$\,nm, and another of $T\sim500$\,K and size $30-50$\,nm \citep{Lau2023ApJ}. Further analysis showed that the ratio of these species varies in the shell number (or rather the age of the dust), and so the dust properties of any CWB are constantly evolving, not only in production, but that composition is also subject to environmental change as the shell ages. 

Another promising avenue to study Wolf-Rayet stars through their spectra is in the method of spectropolarimetry. This method provides an unambiguous probe into the asphericity (or more generally anisotropy) of a stellar wind via any discrepant polarisation between continuum and line emission \citep[][see the latter for a succinct description of the method]{Harries2000A&A, Vink2011A&A}. This method has been successful in identifying wind asphericity and rapid rotation in several classes of stars, including WRs, and even in the WC+O CWB WR\,137 \citep{Harries2000A&A, Lefevre2005MNRAS} although more recent work points to the companion not the WR star as the origin of the asymmetry \citep{StLouis2020}. It is not unambiguously clear, at present, if this method can be made to yield a clear signal indicating stellar rotation in the environment of a WCR.

\subsubsection{Hydrodynamic and Geometric Modelling}
\label{sec:infrared_modelling}
\begin{figure}
    \centering
    \includegraphics[width=0.5\linewidth]{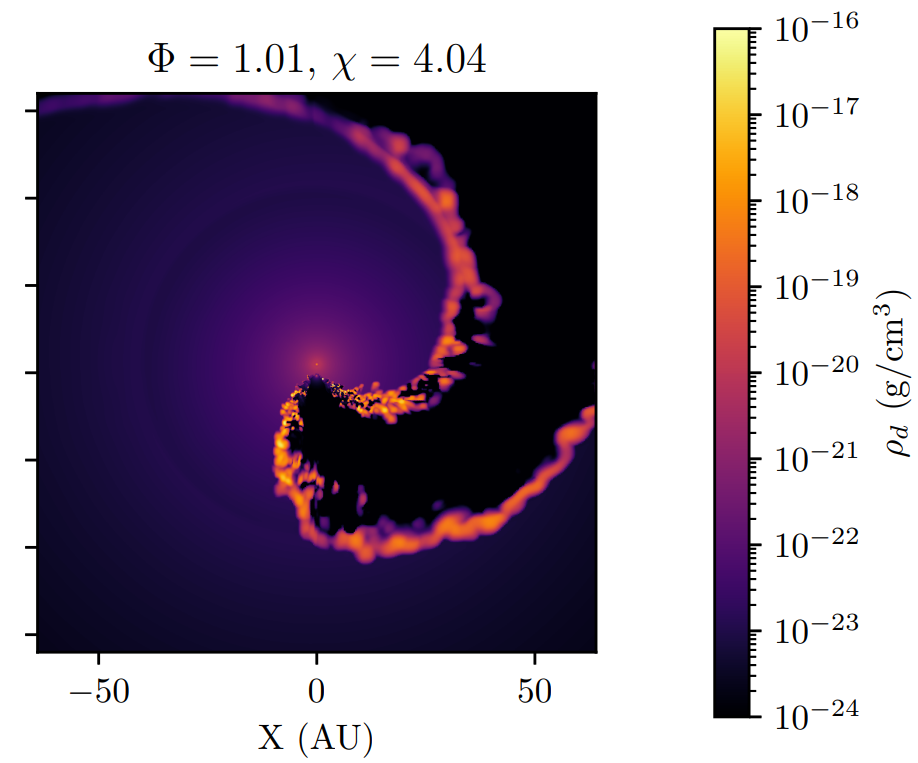}
    \caption{Hydrodynamical simulations can accurately reproduce the conditions that result in dust production along the colliding wind shock. This figure from \citet{Eatson2022MNRASb} shows how the turbulent region in the shocked wind, here just after periastron in WR\,140 where the shock cooling is adiabatic, produces large clumps of dust. The colourbar shows the dust density, $\rho_d$, in the region around the binary.}
    \label{fig:hydro_sim}
\end{figure}

The most rigorous, although also computationally intensive, treatment of the colliding wind environment surrounding Wolf-Rayet binaries is with hydrodynamical simulations. Therein the complicated interactions of gas and dust in the environment around the binary can be simulated, offering insight into dust production and the chemical content of CWB nebulae. It is with these simulations that the dust production mechanism within CWBs was first quantified, showing that gas mixing in the turbulent shocked fluid can efficiently create dust amidst the harsh WR environment (such as in Figure~\ref{fig:hydro_sim}).

These simulations are essential in predicting the geometric dust production around the shock, as well as comparing the state of theory with observations. One of the most easily observed and important parameters describing colliding wind nebulae, no matter the orientation, is the shock opening angle. The opening angle uniquely describes the momentum ratio between the two stellar winds in the binary, defined from stellar mass loss rates and wind terminal velocities as
\begin{equation}
    \eta = \frac{\dot{M}_1 v_{\infty, 1}}{\dot{M}_2 v_{\infty, 2}} \label{eq:windmomentumratio}
\end{equation}
for two stars of subscript 1 and 2. For a range of wind parameters, hydrodynamical simulations have accurately linked the wind momentum ratio and shock opening angle. 
Notably, hydrodynamical simulations were used to suggest that dust production is more efficient around the trailing edge of the shock than the leading edge \citep[with respect to the orbital motion of the stars][]{Lamberts2012A&A, Hendrix2016MNRAS, Soulain2023MNRAS}. This is observationally confirmed by the `azimuthal variation' in dust production around the shock cone of WR\,140 in particular \citep{Williams2009MNRAS, Han2022Natur}. 

Hydrodynamical simulations have helped show just how sensitive dust production is to system parameters, explaining in large part the diversity of colliding wind nebulae. Not only has mass loss rate, wind velocity, and wind chemical composition been shown to alter the shock properties \citep{Eatson2022MNRASa, Soulain2023MNRAS}, but even the speed of orbital motion has been suggested as a key factor in the turn on vs turn off true anomaly in eccentric systems, such as WR\,140 \citep{Eatson2022MNRASb}. Further, these simulations are consistently used in the reproduction of non-thermal (primarily X-ray and radio) observations \citep{Pittard2006MNRAS, Mackey2023MNRAS}.

\begin{figure}
    \centering
    \includegraphics[width=0.8\linewidth]{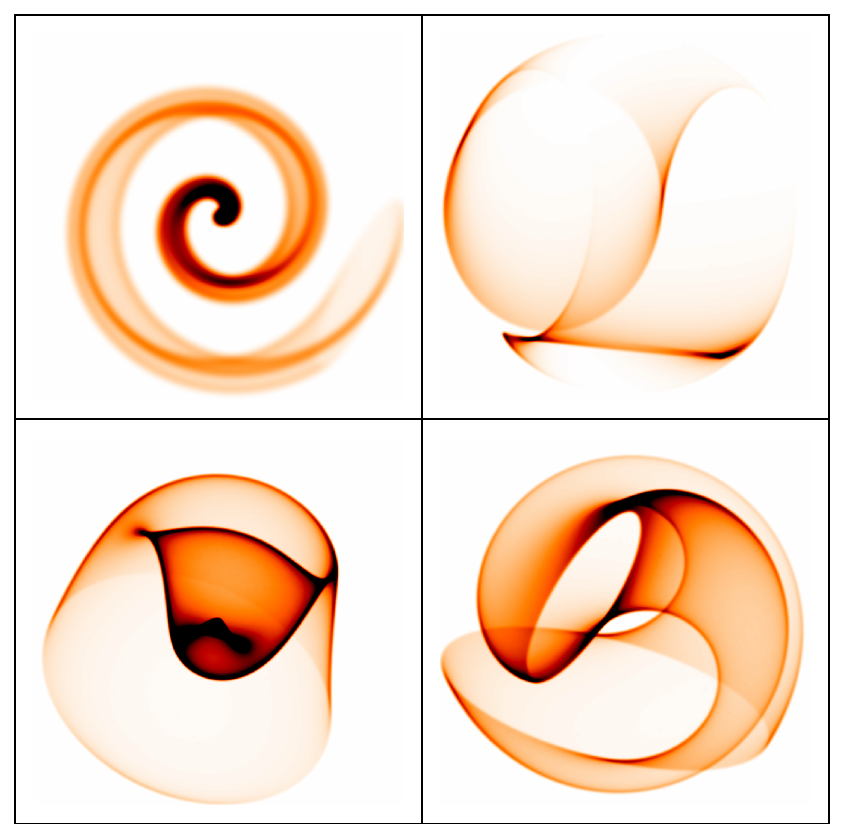}
    \caption{Despite being the result of essentially the same physics, the colliding wind nebulae are able to take on strikingly different apparent forms. From left to right, top down, this mosaic shows simulated nebulae of WR\,104, WR\,140, WR\,112, and Apep using the current best values for wind and orbital parameters in the literature.}
    \label{fig:cwb-mosaic}
\end{figure}

In parallel with the maturation of hydrodynamical simulations, various simple geometric models were developed to reproduce key features of colliding wind nebulae. These geometric models allow for far simpler parameter estimation of CWBs and at much faster compute time compared with hydrodynamical simulations. Archimedean spirals have been used since the first direct images of these dust nebulae \citep{Tuthill1999Natur}, although this does not account for the volumetric geometry of the dust plume and is valid only for a narrow range of binary orbital inclinations and eccentricities. Since then, a more physically motivated geometric model has been developed which reproduces the volumetric structure of the expanding plume (shown in Figure~\ref{fig:cwb-mosaic}). The model approximates the wind collision process by tracking rings of material that expand on the surface of a cone after encountering the nose of the wind-wind shock, expanding with a velocity corresponding to that of the radial wind. When populating a high number of rings over the orbital period, the result is a discretised expanding plume structure that faithfully emulates the true geometry of the dust nebula when interpolated. This family of geometric models was first shown in \citet{Monnier2002ApJ} for WR\,140, but has since been used for Apep \citep{Callingham2019NatAs, Han2020MNRAS}, WR\,104 \citep{Harries2004MNRAS}, WR\,112 \citep{Lau2020ApJb}, WR\,137 \citep{Lau2024ApJ}, and WR\,140 \citep[again, more accurately][]{Han2022Natur, Lau2022NatAs}, with great success when compared to CWB imagery. Such a geometric model was used to show the dust nebula in Figure~\ref{fig:colliding_winds}.

The wind momentum ratio (Equation~\ref{eq:windmomentumratio}) is a useful tool to learn about the relative properties of stellar winds in a binary. Without knowing the mass loss rates and velocities of each stellar wind -- as in what we might see from a direct image snapshot of a colliding wind nebula -- this is difficult to constrain, unlike the observed shock opening angle. Supported by hydrodynamical simulations, two approximate functional forms of the wind momentum ratio are commonly used. For these we need only the wind-wind shock opening angle $\theta_\text{OA}$ of the observed nebula. The first we state is from \citet{Eichler1993ApJ}, 
\begin{equation}
    \frac{\theta_\text{OA}}{2} = 2.1 \left(1 - \frac{\eta_\text{W/S}^{2/5}}{4}\right) \eta_\text{W/S}^{1/3} \label{eq:windmomentumratio_older}
\end{equation}
which when solved -- with $\theta_\text{OA}$ input in radians -- gives information about the momentum ratio of the weaker wind (subscript W) compared to the stronger (subscript S) wind. The second equation we consider is from \citet{Tuthill2008ApJ, Gayley2009ApJ},
\begin{equation}
    \eta_\text{S/W} = \left(\frac{121}{\theta_\text{OA}/2} - \frac{31}{90}\right)^3 \label{eq:windmomentumratio_newer}
\end{equation}
which gives information about the stronger wind relative to the weaker wind. Both of these are approximate forms of the wind momentum ratio, but using them allows one to infer the mass loss properties of the stellar winds using simple geometric modelling only.

Somewhat similar, independent models have been published since the creation of the aforementioned geometric model. The first geometric treatment of the WCR alone was in \citet{Parkin2008MNRAS}, who modelled the shock region as a cone that rotated with orbital motion. This treatment of the geometric model involved implementing a `skew' (a rotation of the WCR ring) to the WCR to account for the orbital motion of the secondary star. While this is apparently necessary to accurately model non-thermal emission from CWBs, it is not clear if this has a significant effect on dust production that would manifest in infrared imagery. An independent attempt at modelling the entire volumetric structure of the dust nebula was published in \citet{Williams2009MNRAS} for WR\,140, where the resulting eccentric structure is close to that of \citet{Monnier2002ApJ}, but with added effects of dust asymmetry not unlike \citet{Han2022Natur}.

These simpler geometric models offer significant computational efficiency at a cost of physical insight. Where a hydrodynamical simulation may require runtime of hours, days, or even more, the geometric model can be evaluated on the order of seconds. This opens the door for parameter estimation using the geometric model rather than using it purely as a consistency check. As mentioned earlier, this does come at a cost of physical insight; these geometric models do not offer detailed information about dust production (e.g. dust mass, non-thermal emission, etc.), but they do offer immediate information on fundamental parameters such as orbital characteristics and dust turn on/off. As of yet, geometric models have not been used to constrain orbital parameters (eccentricity, inclination, etc.) effectively, and a proper statistical treatment with Bayesian parameter inference methods has not yet appeared in the published literature.

\subsection{High-Energy Emission}
Colliding wind binaries are strong X-ray sources, and were first hypothesised as such in \citet{Cherepashchuk1976SvAL} and \citet{Prilutskii1976SvA}. The intrinsic energetics of the colliding winds can boost these emissions to be significantly stronger than those from individual stars in isolation. Since the association of CWBs with X-ray sources, the theory modelling their properties has matured and accurately represents observations \citep[][for a review]{Rauw2016AdSpR}. Modern models reproduce X-ray spectra exceedingly well, providing insight into the mass loss rates and wind speed but also the homogeneity of the wind and its cooling physics \citep{Zhekov2021MNRAS, Zhekov2022MNRAS}. The X-ray flux of eccentric CWBs is variable as the properties at the WCR change over the course of the orbit, allowing us to constrain orbital and dust production properties; the phase of the flux peaks has been found to be wavelength dependent as the most efficient cooling mechanism at the WCR changes between radiative and adiabatic with orbital phase \citep{Gosset2016A&A, Rauw2016AdSpR, Mackey2023MNRAS}. This kind of variation within the light curve is in conjunction with the usual eccentricity effects, as well as a WCR column density constraint on the orbital inclination in the X-ray \citep[see Section 5.3 of][]{Gosset2016A&A}. At some inclinations, the observed X-ray emission strength may as well vary in time over the binary orbit as the nose of the wind-shock passes behind, and is partially obscured by the dense WR wind, only to rise again at a later orbital phase when traversing the thinner wind of the O star. These patterns are observed, for example, in the bright WR binary $\gamma^2$~Velorum \citep{Stevens1995Ap&SS}.

Given the inherent energy associated with the colliding stellar winds, we might also expect that these systems are highly luminous $\gamma$-ray emitters. Indeed, hydrodynamical modelling of colliding winds suggests that there should be a significant $\gamma$-ray emission \citep{Huber2021A&A, Pittard2020MNRAS}. Despite this, only two massive colliding wind binaries have been observed with such a $\gamma$-ray excess: $\eta$~Carinae \citep[consisting of a LBV+? binary;][]{Reitberger2015A&A} and $\gamma^2$~Velorum \citep[a WC8+O7.5 system, also known as WR\,11;][]{North2007MNRAS, Pshirkov2016MNRAS}. Why $\gamma$-ray emission is not abundantly visible in other CWB systems is a topic of active research. The likely reason for this lies in the WCR itself: the turbulent shocked flow is understood to be highly magnetised with a magnetic differential going from the WR side of the shock to the companion side. \citet{Grimaldo2019ApJ} find that proton acceleration in the WCR reduces this field differential and hence reduces the maximum energy of their relativistic motion, in turn hindering their eventual gamma ray emission. 

The same processes that produce high energy photons in these CWBs should also accelerate electrons and protons to relativistic velocities. \citet{DeBecker2013A&A} share the first catalogue of CWBs that efficiently accelerate particles to such energies and briefly review the topic. The majority of the catalogue is composed of Wolf-Rayet binaries with high radio luminosities. In general, the highest energy particles accelerated in these systems will be protons, and with maximal energies of $\gtrsim 1$\,TeV \citep{Reitberger2014ApJ}. The reason for this lies in the cooling mechanism at the colliding wind shock itself. As a rough measure of whether the WCR region is cooling adiabatically or radiatively, \citet{Stevens1992ApJ} introduced the cooling parameter,
\begin{equation}
    \chi = \frac{v_8^4 d_{12}}{\dot{M}_{-7}}
\end{equation}
which when $\ll 1$ is indicative of radiative cooling and adiabatic otherwise. Here, $v_8$ is the wind velocity in units of 1000\,km\,s$^{-1}$, $d_{12}$ is the star-WCR distance in units of $10^7$\,km, and $\dot{M}_{-7}$ is the stellar mass loss rate in units of $10^{-7}\,M_\odot$\,yr$^{-1}$. In the time since that paper, however, \citet{Mackey2023MNRAS} have introduced an analogous parameter,
\begin{equation}
    \chi_\text{IC} = \frac{1.61 v_8 d_{12}}{L_5}
\end{equation}
which accounts for the effect of inverse-Compton cooling in the shocked wind, or cooling by additional optical line emission that briefly supplants X-ray cooling \citep{Pollock2021}. Here, $L_5$ is the luminosity of the star in units of $10^5 L_\odot$. Again, $\chi_\text{IC} < 1$ is evidence towards a radiatively cooling wind. Understanding whether the shocked material in the WCR is cooling radiatively or adiabatically in turn informs us of the efficiency high energy emission and particle acceleration. When the cooling is radiative, the plasma efficiently dissipates energy and free particles do not become as energetic \citep{Reitberger2014ApJ, Rauw2016AdSpR}. Conversely, adiabatic cooling typically results in accelerated electrons maintaining their energy. Protons are not affected by the same cooling mechanisms as electrons and in fact retain their energy within the WCR; this is why accelerated protons are more readily produced in CWBs, especially where the shock is radiatively cooling \citep{Grimaldo2019ApJ}.

\subsection{Radio}
\begin{figure}
    \centering
    \includegraphics[width=0.6\textwidth]{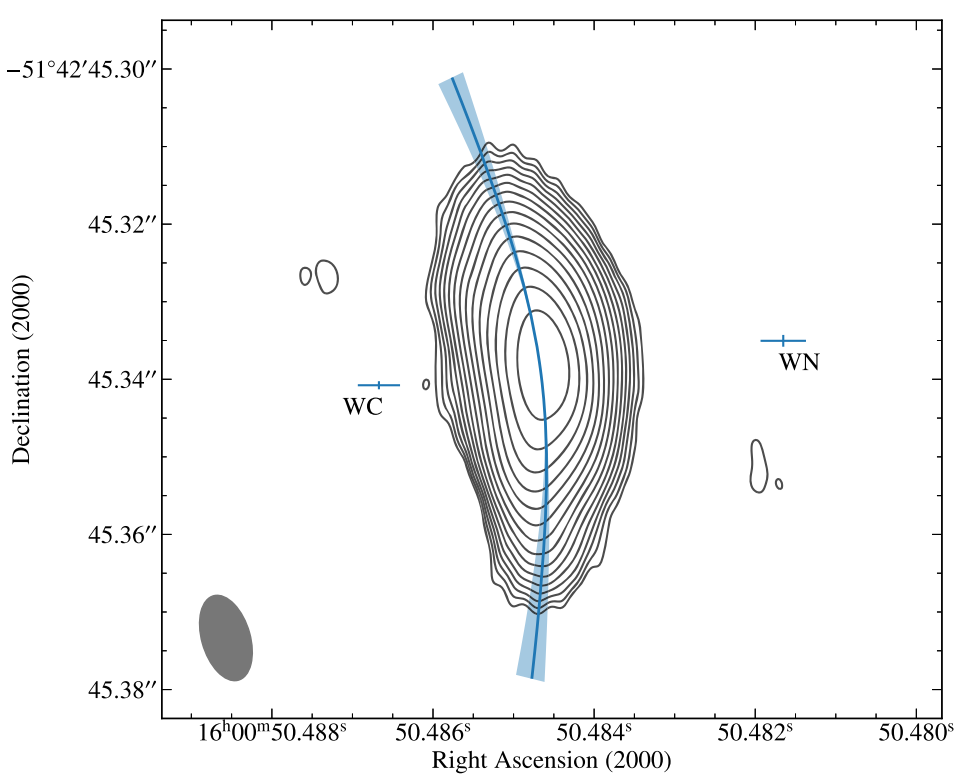}
    \caption{The wind collision region can be directly imaged in radio wavelengths. In this figure from \citet{Marcote2021MNRAS}, we see the non-thermal emission at the bow shock between the two stars of the Apep system; the blue curve overlay shows the ideal shock boundary for a $\eta = 0.44 \pm 0.08$ wind collision.}
    \label{fig:apep_radio}
\end{figure}
Colliding wind binaries can be extremely luminous in the radio and yield a rich phenomenology amenable to study with modern facilities as well as a wealth of historical data stretching as far back as 1976 \citep{Seaquist1976ApJ, DeBecker2013A&A}. In the radio, there are two types of physical processes that generate observed emissions. One is thermal emission due to free-free radiation in which charged particles (free electrons and ions) in the stellar wind rapidly accelerate under collisional electrostatic forces, emitting radiation while quickly reaching local thermal equilibrium \citep{Wright1975MNRAS}. The second source is non-thermal emission from synchrotron radiation, in which ultra-relativistic charged particles such as free electrons emit radiation in the radio while spiralling around magnetic field lines \citep{Condon1992ARA&A, Eichler1993ApJ, Pittard2006A&A}. These two emission mechanisms may co-exist in a single system, and even be spatially resolved at different locations as is the case of WR147, where the thermal emission is shown to originate closer to the WN star with the stronger wind, while the non-thermal emission is produced closer to the B companion \citep{Williams1997MNRAS}.

The non-thermal emission from the colliding wind binaries, and its variability, can inform us of many physical phenomena at work. This non-thermal emission is particularly susceptible to absorption effects which can hinder the radio observability of these systems. Many absorption mechanisms work in parallel and to different strengths depending on the conditions at the wind collision region. Where the WCR is especially dense (for example when the two stars are close together, $\lesssim 10$\,au, in their orbits), free-free absorption \citep{Benaglia2020PASA, Blanco2024A&A} and synchrotron self-absorption \citep{Kellermann1966AuJPh, Callingham2015ApJ} work to decrease the escaped flux at low frequencies. Similarly, the Razin-Tsytovich effect also reduces the radio flux in regions of highly magnetised material in the WCR \citep{DeBecker2007A&ARv}. All of these mechanisms together paint a picture where low energy ($\lesssim 1$\,GHz) non-thermal emission is heavily modulated by the orbit of the stars, and wherein orbital progression results in strong variation for sightlines passing through the intervening ionised winds which can render them opaque to the inner WCR. Conversely, higher energy radio photons are not as affected by such absorption processes and show comparatively less modulation with the orbital phase. 

Radio observations of CWBs offer information of the WCR directly, even so far as identifying the WCR itself with interferometric imaging \citep[][Figure~\ref{fig:apep_radio}]{Dougherty2005ApJ, Rodriguez2020ApJ, Marcote2021MNRAS}. Since radio observations constrain the WCR so well, any discrepancies between wind models and data (e.g. momentum ratios, mass loss rates, etc.) are immediately apparent.

\subsection{A selection of notable galactic WR CWBs} \label{sec:catalogue}
\begin{figure}
    \centering
    \includegraphics[height=.32\columnwidth]{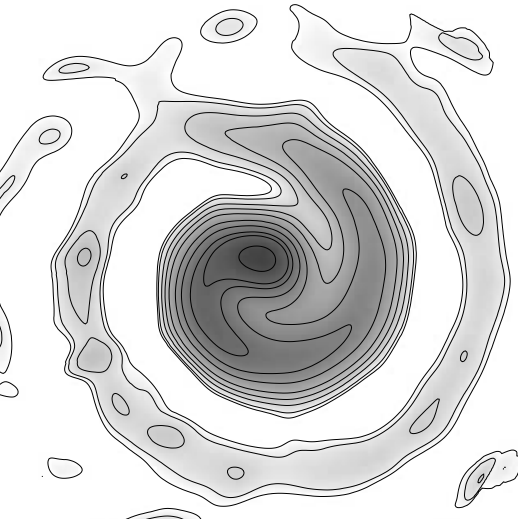}
    \includegraphics[height=.32\columnwidth]{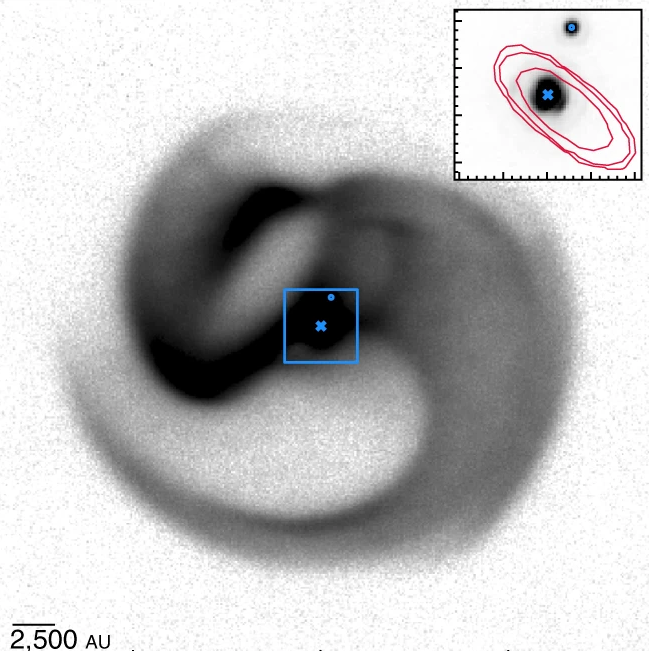}
    \includegraphics[height=.32\columnwidth]{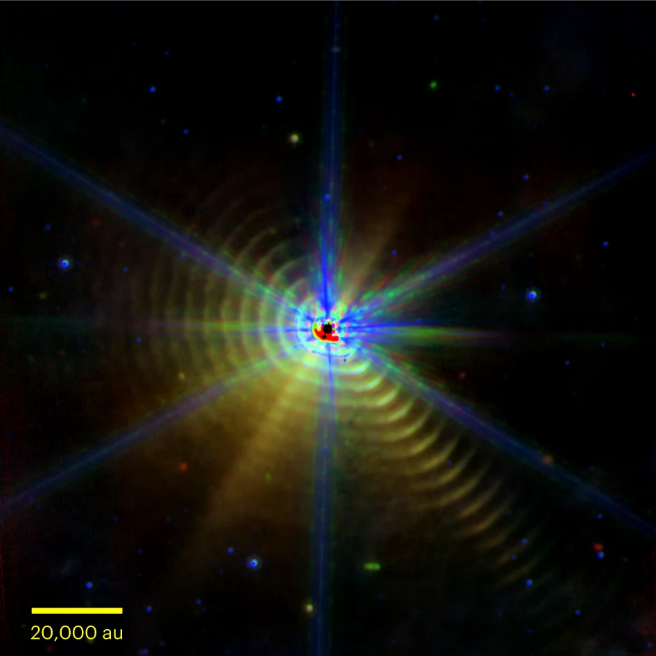}
    \caption[Images of WR\,104, Apep, and WR\,140]{The last $\sim20$ years has seen much progress in imaging the dust nebula colliding wind binaries. \emph{Left:} The first imaged CWB Pinwheel, WR\,104 \citep{Tuthill2008ApJ} recovered from  interferometric data. \emph{Centre:} The only confirmed WC+WN system in the Galaxy, Apep \citep{Callingham2019NatAs}, taken with the VISIR instrument on the Very Large Telescope (VLT). \emph{Right:} The first CWB directly imaged with the \emph{James Webb Space Telescope}, WR\,140 \citep{Lau2022NatAs}, bears witness to more than a century of dust production in the form of $\sim$20 concentric nested shells.}
    \label{fig:CWBs}
\end{figure}
This section presents some of the most unique and significant (historically or otherwise) Wolf-Rayet colliding wind binaries for ease of reference with an emphasis on systems for which rich studies across the spectrum promise to yield a more complete picture of the astrophysics. 
\begin{description}
    \item[$\gamma^2$ Velorum] As a WR+O system, itself part of a higher order 4-star system, $\gamma^2$ Velorum (WR\,11) is one of the most constrained CWB orbits yet, with  \citet{North2007MNRAS} providing a full orbital solution from astrometry that is orders of magnitude more precise than other methods. This system is the closest and brightest WR system at merely $336^{+8}_{-7}$\,pc \citep{Millour2007A&A, North2007MNRAS}. Despite hosting a WC8 star \citep{DeMarco1999A&A, DeMarco2000A&A}, the $\gamma^2$ Velorum system does not appreciably produce the dust that is typical of these binary configurations \citep{van_der_Hucht1996A&A}. With that said, the spiral density perturbations arising from the WCR (with analogous phenomenology to the dust nebulae) have been used to explain the `enshrouded' companion O star spectrum in the system \citep{Lamberts2017MNRAS}.

    \item[WR\,19] This system is one of the most eccentric and longest period colliding wind binaries known, with $e \simeq 0.8$ and $P_\text{orb} = 10.1\pm0.1$\,yr \citep{Veen1998A&A, Williams2009MNRASb, Williams2019MNRAS}. At present, there are no direct images of WR\,19 published in the literature, although it is revealed as a site of strong dust formation through its infrared photometry and spectra. Despite having no image, the period and eccentricity of the orbit suggests that it may have similar structure to the nebulae of WR\,140. 
    
    \item[WR\,48a] At present, this is the only other candidate WR+WR CWB -- specifically of WC8+WN8h classification \citep{Zhekov2014MNRAS} -- and found in an orbit with eccentricity $e\sim0.6$ \citep{Williams2012MNRAS}. This was among the first objects to be identified as a colliding wind binary \citep{Danks1983A&A}, and was first directly imaged in \citet{Marchenko2007ASPC}. Like Apep, WR\,48a has a relatively long orbital period of $\sim32$ years and displays variable dust production throughout its orbit (although there is no evidence of a complete dust `turn off' in 48a analogous to that in Apep).

    \item[Apep] One of the only confirmed WR+WR systems, and certainly the only such to harbour a WC with a dust nebula \citep{Callingham2020MNRAS}, Apep (WR~70-16) was identified as a bright X-ray and radio source and is potentially the brightest of all the CWBs. It has been observed with the VLT in the mid-infrared which revealed its spiral dust nebula structure \citep{Callingham2019NatAs, Han2020MNRAS}. Also a likely triple system, an O8 star lies very close (0.7") to the north of the WR+WR inner binary with evidence that indicates physical association.
    
    \item[WR\,98a] One of the first colliding wind nebulae to be directly imaged, this system showed the first imaging evidence of inclination effects in a pinwheel nebula \citep{Monnier1999ApJ}, in contrast to the Archimedean spiral of WR\,104. 
    
    \item[WR\,104] As the first colliding wind binary to be directly imaged at high angular resolution, WR\,104 emerged as the prototype `pinwheel nebula' from the clear evidence of orbital motion in its dust plume \citep{Tuthill1999Natur}. This system has been the subject of intense study, particularly in hydrodynamical simulations and the application of pinwheel structure to other CWB nebulae. Like WR\,98a, WR\,104 is a reasonably short period WC+O binary ($\lesssim2$ years) in a very circular orbit \citep{Tuthill2008ApJ}. Recently, a B star distant from the WC+O binary was considered as being associated with the system \citep{Wallace2002ASPC, Soulain2018A&A}, potentially making WR\,104 one of the very few known WR triple systems. \citet{Hill2024MNRAS} suggested that the binary orbit may in fact be moderately inclined from our perspective (inclination $i>34^\circ$: a finding that, at face value, is difficult to reconcile with the face-on fits $i<16^\circ$ from imagery. Any resolution to this puzzle remains outstanding at the time of writing.  
    
    \item[WR\,112] First imaged in 2002, the colliding wind nebula of WR\,112 was the first to visibly reveal concentric shells of dust formation \citep{Marchenko2002ApJ} from its inner WC8+O binary \citep{Lau2020ApJb}. Like the other consistent dust producers listed here (WR\,98a, 104), WR\,112 has a well constrained eccentricity close to 0 as is evident from its regular concentric shell structure. Notably, the pinwheel nebula structure was incorrectly applied to WR\,112's dust nebula twice, yielding far-off estimates of the nebula expansion speed and the binary eccentricity and inclination \citep{Marchenko2002ApJ, Lau2017ApJ}.

    \item[WR\,125] One of the longer period colliding wind binaries, the WC7+O9III binary shows episodic dust production with a periodicity of 28.1 years \citep{Endo2022ApJ, Richardson2024arXiv}. While not yet imaged, the surrounding dust shell has been suggested as anisotropic and the binary orbit elliptical \citep{Richardson2024arXiv}.

    \item[WR\,137] One of the three original systems first catalogued by French astronomers Charles Wolf and Georges Rayet in 1867 \citep{Wolf1867CRAS}, WR\,137 is a WC7+O9V colliding wind binary \citep{Richardson2016MNRAS}. The infrared light curve shows episodic dust production over an orbital period of about 13.1 years, and the orbital plane of the binary appears to be almost entirely in the line of sight \citep{Peatt2023ApJ, Lau2024ApJ}.
    
    \item[WR\,140] The `prototype' colliding wind binary, composed of WC8+O5 stars on a very high eccentricity ($e \sim 0.89$) orbit of about 7.9 years \citep{Williams2009MNRAS, Monnier2011ApJ}. This was the first system to be unequivocally identified as a colliding wind binary, largely due to its strongly modulated dust production: the dust seen in Figure~\ref{fig:CWBs} (right-side) is produced only in a narrow time frame as the stars approach and leave periastron \citep{Lau2022NatAs, Han2022Natur}, and this reliably repeats on each orbital passage. This is perhaps the most well studied of all CWBs across all wavelengths and imaging techniques. 
    
    \item[WR\,147] An apparent triple system composed of a WN8, a close unseen companion, and a distant OB star \citep{Rodriguez2020ApJ}. This is the first system to have an observed pinwheel from the inner binary not in the infrared but in the radio. Furthermore there is a well resolved bow-shock distinct from the pinwheel where the winds collide with the ternary component. This aligns with previous X-ray observations that resolved the system as a double X-ray source: one at the inner binary and another at the WCR of the inner binary relative to the outer companion \citep{Zhekov2010ApJa, Zhekov2010ApJb}.

\end{description}

\section{Implications and Fate}

\subsection{Wolf-Rayet Binaries as Supernova Progenitors}

Wolf-Rayet stars are convincing candidates for Type Ib/Ic supernovae (SNe~Ibc) progenitors: WRs are massive pre-supernova stars that lack hydrogen spectra, as do SNe~Ibc \citep{Filippenko1997ARA&A, Crowther2007ARA&A, McClelland2016MNRAS}. There is some discussion of an alternate SNe~Ibc channel in \citet{Smartt2009ARA&A} where a lower mass star may lose its hydrogen envelope via binary interactions prior to undergoing supernova \citep[not dissimilar to one of the proposed channels of WR formation;][]{Paczynski1967AcA}. The consensus is that at least some combination of the WR progenitor model and this alternate scenario explain the observed population of SNe~Ibc \citep[see:][]{Eldridge2013MNRAS, Pellegrino2022ApJ, Karamehmetoglu2023A&A}. 

Despite their intuitive connection, no WR stars have been definitively shown to be SNe~Ibc progenitors with pre-supernova imaging of galaxies. \citet{Kilpatrick2021MNRAS} identified the likely progenitor of Type Ib supernova SN2019yvr in \emph{Hubble Space Telescope} imagery taken 2.6\,yr prior to explosion. They found that neither a clear-cut WR nor a lower mass binary can adequately describe the SN progenitor, indicating that perhaps a sudden change in the observables of SNe~Ibc progenitors take place just prior to explosion, e.g. extreme mass loss. The lack of attribution could also plausibly be ascribed at least in part to the reduced detectability of WR stars at extragalactic distance \citep{Pledger2021MNRAS}. 

On the other hand, connections have been made between Type Ic supernovae and gamma-ray bursts (GRBs). Gamma-ray bursts are a class of energetic transient whose durations distinctly fall into two categories of `short' and `long', with several sources appearing isotropically on the sky each day \citep{Woosley2006ARA&A}. Short GRBs, of order $\lesssim 2$\,s, are understood to be the result of kilonovae, the breakup of degenerate neutron stars during gravitational wave induced binary inspiral \citep{Berger2014ARA&A}. In contrast, long GRBs (LGRBs), of order $\gtrsim 2$\,s, arise from the relativistic jets produced in the core-collapse of highly rotating massive stars \citep{Woosley2006ApJ}. In fact, several SNe~Ic-BL events, SNe~Ic with especially broad spectral lines indicative of relativistic ejecta, have been definitively associated with LGRBs and fast X-ray bursts \citep{Galama1998Natur, Ashall2019MNRAS, van-Dalen2024arXiv}. Interestingly, not all of these SNe~Ic-BL have been associated with LGRBs, indicating that either the jets are highly collimated (and hence their visibility being very direction dependent) or there remains a significant gap in understanding \citep{Siebert2024arXiv, van-Dalen2024arXiv}.

The consensus on the LGRB mechanism is in the collapsar model, where at the time of supernova the innermost core collapses into a black hole (BH). The surrounding core, having a high enough angular momentum to form a disc, produces polar jets as a result of angular momentum loss as matter accretes onto the natal BH \citep{Woosley1993ApJ, Dean2024PhRvD}. For a star to do this, it must be rotating extremely rapidly. This should be the case for stars that have been efficiently mixed during their lifetimes \citep{Woosley2006ApJ}, and hence stars that are effectively free of hydrogen at supernova as a result. Therefore it is not a requirement that GRBs have no hydrogen line association (and hence a Type I vs a Type II SN), but rather a consequence of the evolutionary process that gives rise to GRBs. Because of this, helium and metal-rich stars emerge as the leading candidate progenitors for SNe~Ic/-BL. There is a significant body of simulations linking WR to SNe~Ic and GRBs through the collapsar model \citep{McClelland2016MNRAS, Aguilera-Dena2018ApJ}, especially those with high rotation at preferentially low metallicity \citep{Detmers2008A&A}.

\subsubsection{Metallicity in Different Epochs}
Modern Wolf-Rayet stars are synonymous with metallic line-driven winds. These winds, along with magnetic effects, are known to be efficient at dissipating angular momentum from hot stars \citep{Woosley2006ApJ, Ud-Doula2009MNRAS}. This, in conjunction with the more efficient mass-loss at high metallicity \citep{Vink2005A&A, Grafener2008A&A}, means that we generally expect Galactic WRs to be slower rotators than those in less-evolved galaxies. There is a strong dependence on the rotation speed of a progenitor star to its LGRB status \citep[and also between the spin of the collapsar BH to the GRB jet collimation;][]{Hurtado2024ApJ}, and so the logical conclusion is that we should expect fewer or no LGRBs in evolved galaxies with high metal content. 

Indeed, there is strong indication from stellar models that low-metallicity massive stars should be rapidly rotating \citep{Chiappini2006A&A, Vink2017A&A}. This motivates searches for LGRB progenitors in dwarf and/or satellite galaxies where the metal content can be orders of magnitude below solar metallicity \citep{Tolstoy2009ARA&A}. There has been some plausible, if inconclusive, evidence of rapidly-rotating WRs in the Large Magellanic Cloud \citep{Shenar2014A&A, Vink2017A&A}, lending credence to this idea. Therefore we expect that LGRBs are efficiently produced in the early universe or in dwarf galaxies with less enriched chemical content. 

One proposed channel for fast spinning massive stars, and hence LGRB events, within chemically evolved galaxies like the Milky Way is in the tidal spin up from a closely orbiting companion star. Early studies suggest that this mechanism is improbable at solar metallicity and marginally more likely at low metallicity \citep{Detmers2008A&A}. More recently, population synthesis studies have shown this to be an efficient mechanism to sustain fast rotation in evolved stars \citep{Chrimes2020MNRAS, Bavera2022A&A}, although only in very closely orbiting binaries. While most observed evolved Wolf-Rayet stars are not in close binaries, a select few (namely WR\,140 and Apep) are in highly eccentric orbits which brings the two component stars together at periastron. It is unclear how these close passages affect the rotation and internal structure of the WR stars, and if this is allows for a tidal spin-up.

\subsubsection{Supernovae in Binaries and Effects of Circumstellar Material}
Regardless of the stellar rotation properties, a significant fraction of Wolf-Rayet stars exist in binaries and their supernova explosions are astrophysically imminent. If a Wolf-Rayet supernova occurs in a colliding wind binary system, especially one with a prominent infrared excess, there will be a rich circumstellar structure surrounding the system from previous epochs of dust production. After supernova begins and light begins emanating from its source, the surrounding material may become luminous (with some delay from the explosion) by the `light echo' phenomenon \citep{van_den_Bergh1965PASP, Crotts1995ApJ}. 

The radial extent of the colliding wind nebulae discussed earlier are determined mainly by two parameters: the expansion velocity and the orbital period. In this way, any light echo inherent to this circumstellar nebula structure would have an emission delay proportional to each of these parameters and the speed of light. A consequence of this periodic shell circumstellar medium (CSM) structure is that the supernova light curve should also display some periodicity in its light echo (or in the repeated SN ejecta-CSM contact). \citet{Moore2023ApJ} report the first unambiguous detection of such a light curve in the optical, with a $\sim 12.5$~day modulation of brightness. This supernova, SN\,2022jli, was also of the Type~Ic subtype indicative of a Wolf-Rayet progenitor; further, the required CSM mass is consistent with typical WR mass loss rates. Similar quasi-periodic bumps in brightness have been observed in SN\,2015bn \citep{Nicholl2016ApJ}, SN\,2020qlb \citep{West2023A&A}, and SN\,2023aew \citep{Kangas2024A&A}, among others, which are all hydrogen-poor SNe albeit with higher time intervals between light curve bumps than in SN\,2022jli. 

Turning away from the optical emission from supernovae and into the radio, unambiguous periodic oscillations in brightness have been observed twice before: in SN\,1979C \citep{Weiler1991ApJ, Weiler1992ApJ} and in SN\,2001ig \citep{Ryder2004MNRAS}. As of yet, there has been no definite periodic radio emission observed around a SNe~Ibc, although SN\,2007bg likely shows evidence of a structured CSM through it's radio emission \citep{Salas2013MNRAS}. No matter the supernova subtype, there is evidence across the electromagnetic spectrum indicating a structured circumstellar medium surrounding massive, evolved stars. While ejecta interactions with stellar mass eruptions may support a subset of these observations, the nested rungs of regular dust shells created from colliding wind binaries provide the most natural explanation for these environments.

\section{Summary and Outlook}
Colliding wind binaries, especially those hosting a Wolf-Rayet star, open a new window into otherwise inaccessible stellar physics. The wind-wind shock between the stars allows us to better understand the properties of each stellar wind, and in turn offers a rich landscape of observable phenomenology that yields diagnostics and probes into the final phases of massive stars immediately before undergoing supernova explosions. These wind collision shocks are bright non-thermal emitters in the X-ray and at radio frequencies, and can be luminous in the infrared and sub-millimetre in cases where the production of copious warm dust is nucleated from gas in the streaming wind. With many examples offering a panoply of intricately structured observational signatures, the Wolf-Rayet CWBs offer a fascinating astrophysical laboratory open to studies across the electromagnetic spectrum. 

Every year, more Wolf-Rayet colliding wind binaries are discovered in the Galaxy, the Magellanic Clouds and increasingly at intergalactic distances. As higher cadence and wider field photometric surveys come online, our understanding of the spatial distribution of CWBs as well as their time-evolving behaviour will come under close observational scrutiny. In parallel, sensitive infrared and sub-millimetre observatories such as JWST and ALMA offer a new avenue to progress our understanding of the orbits of these stars and the geometries of their nebulae. Although the separation of the binary stars themselves is usually too close and/or the system too obscured to be directly resolved, the fast wind delivers a remarkable gift to observers. Complex structures encoding detailed colliding-wind physics, engraved at sub-milliarcsecond scales, are preserved and inflated within the spherically expanding nebula, leaving a fossil record witnessing hundreds of years of system evolution written across the sky at arcsecond scales. While contemporary work has delivered remarkable success in fitting these data with simple geometrical models, a synthesis with physically motivated elements from hydrodynamics and radiative transfer will be required to understand upcoming and archival direct imagery. 

Given the disproportionate importance played in the cosmic ecosystem by the most massive of stars -- which drive radiation, winds and supernova ejecta at galactic scales -- astronomy is fortunate to have the colliding wind binaries.
Each forms a distinct and richly featured laboratory boasting diagnostics accessible across the electromagnetic spectrum for confronting model scenarios, constraining basic properties and yielding insight into this fleeting yet critical phase in the lives of massive stars.

\section*{Acknowledgements}
We thank Benjamin~Pope, Yinuo~Han and Peredur Williams for helpful discussions, suggestions, and feedback throughout the writing of this work. RMTW acknowledges the financial support of the Andy Thomas Space Foundation.

% \setcitestyle{numbers} % uncomment this to see the number of references
% % \bibliographystyle{elsarticle-harv} 
% \bibliographystyle{mnras} % just using this in the meantime for ads links
% % \bibliography{cas-refs, references}
\bibliographystyle{Harvard}
\bibliography{references}





% \end{thebibliography}
\end{document}